\newif\iftwocolumns
\twocolumnsfalse     

\iftwocolumns
  \documentclass[aps,prl,twocolumn,groupedaddress,amsmath,amssymb,showpacs]{revtex4}
\else
  \documentclass[aps,prl,preprint,groupedaddress,amsmath,amssymb,showpacs]{revtex4}
\fi

\usepackage{graphicx}
\usepackage{dcolumn}
\usepackage{bm}

\def\const{\mathop {\rm const}\nolimits } 
\begin{document}



\title{Force and moment balance equations for geometric variational problems on curves}


\author{E.~L.~Starostin}
\email{e.starostin@ucl.ac.uk}
\email{eugene.starostin@daad-alumni.de}
\author{G.~H.~M.~van~der~Heijden}
\email{g.heijden@ucl.ac.uk}
\affiliation{Centre for Nonlinear Dynamics, University College London,\\
 Gower Street, London WC1E 6BT, UK}

\date{\today}

\begin{abstract}
We consider geometric variational problems for a functional defined on a
curve in three-dimensional space. The functional is assumed to be written in
a form invariant under the group of Euclidean motions. We present the
Euler-Lagrange equations as equilibrium equations for the internal force and
moment. Classical as well as new examples are discussed to illustrate our
approach. This new form of the equations particularly serves to promote the
study of bio- and nanofilaments.
\end{abstract}


\maketitle


\section{Introduction}

Ever improving experimental techniques in biophysics and nanotechnology
have created great interest in
one-dimensional continuum models for such slender structures as DNA,
proteins, nanotubes and other bio- an nanofilaments
\cite{Helfrich91,Feoli05,Tu08,Starostin08a}. In addition, such
models continue to be used in engineerinig applications to study large
statical deformations of one-dimensional elastic structures (e.g., cables,
pipelines, textile yarns) \cite{Heijden01,Starostin07}.
Vortex filaments provide another target for the application of 
one-dimensional continuum models \cite{Langer96,Barros07}.

Often these models give rise to variational problems on
curves in a form that is invariant under Euclidean motions. The corresponding
(Euler-Lagrange) equilibrium equations are usually derived ad hoc. To be
sure, there is a general theory of Euler-Lagrange equations for invariant
variational problems \cite{Anderson89,Kogan03}, but it is usually expressed
in abstract geometrical form and does not seem to be widely known in the
physics and mechanics literature. Moreover, the equations it yields are
naturally expressed as high-order ordinary differential equations (ODEs),
which are neither necessarily convenient for further analysis or numerical
solution, nor helpful in providing insight into the problem under
consideration.

Here we show that Anderson's Euler-Lagrange equations of
\cite{Anderson89} for variational problems on curves can be written in the
form of (first-order) balance equations for the internal forces and moments
in the structure plus equations that can be interpreted as constitutive
relations. We believe that this form of the equations is better suited to
further analysis, in particular in problems of rods and filaments
subjected to end loads (as, e.g., in single-molecule experiments). We
demonstrate the usefulness and wide applicability of the equations by a
series of classical as well as novel variational problems.

\section{Equations for invariant variational problems}

Consider the variational problem for a functional $f$ on a smooth curve
${\cal C} =\{\bm{r}(s) \in \mathbb{R}^3, s \in [0,L]\}$:
\begin{equation}
\int_0^L f[s,\bm{r}(s),\chi(s)]\,\mbox{d}s \to \mbox{extr}.
\label{eq:func0}
\end{equation}
Here $\chi(s) \in \mathbb{R}^n$ collects possible additional functions
defined on the curve. We assume that the scalar function $f$ is invariant
under 
reparametrisations of the curve $\cal{C}$ and invariant under the group of
Euclidean motions of $\mathbb{R}^3$. Then the functional can be expressed in
terms of the Euclidean invariant properties of the curve $\cal{C}$, i.e.,
its curvature and torsion. For such problems it is possible to write down the
Euler-Lagrange equations directly in terms of the geometric invariants, i.e.,
avoiding coordinates $\bm{r}$ \citep{Kogan03}. The following proposition
(first briefly announced in \citep{Starostin07a} in a slightly less general
form) gives the equations in the form of force and moment balance equations.
The result is a natural convergence of lines of work in mechanics, physics
and mathematics that can be traced back to
\citet{Sadowsky31},
\citet{Langer96},
\citet{Capovilla02} and, in more abstract form, to the theory of the
invariant variational bicomplex
\citep{Griffiths83,Anderson89,Kogan03}.

{\it Proposition.} 
Let $\bm{r}(s), s \in [0,L]$, be a sufficiently smooth regular curve in $\mathbb{R}^3$ with unit tangent vector $\bm{r}'(s)=\bm{t}$, curvature $\varkappa(s)$ and torsion $\tau(s)$. Here and in the following the prime denotes differentiation with respect to arclength $s$. In addition, let $\chi(s)$ be a smooth function of arclength.
Then the Euler-Lagrange equations for the variational problem
\iftwocolumns
\begin{eqnarray}
\int\limits_0^L f(\varkappa,\tau,\chi,\varkappa',\tau',\chi',\varkappa'',\tau'',\chi'',\ldots,\varkappa^{(p)},\tau^{(q)},\,\chi^{(r)})\,\mbox{d}s \to 
\nonumber \\
\to \mbox{extr} 
\label{eq:funcl}
\end{eqnarray}
\else
\begin{equation}
\int_0^L f(\varkappa,\tau,\chi,\varkappa',\tau',\chi',\varkappa'',\tau'',\chi'',\ldots,\varkappa^{(p)},\tau^{(q)},\,\chi^{(r)})\,\mbox{d}s \to \mbox{extr} 
\label{eq:funcl}
\end{equation}
\fi
can be presented in the form of (a) balance equations for the components of
the internal force $\mathsf{F}=(F_t,F_n,F_b)^T$ and moment
$\mathsf{M}=(M_t,M_n,M_b)^T$ expressed in the Frenet frame $\{\bm{t},\bm{n},\bm{b}\}$ (tangent, principal normal, binormal),
\begin{equation}
\mathsf{F}'+\mathsf{\omega}\times\mathsf{F}=\mathsf{0}, \quad\quad \mathsf{M}'+\mathsf{\omega}\times\mathsf{M}+\mathsf{t}\times\mathsf{F}=\mathsf{0},
\label{eq:balance}
\end{equation}
where $\mathsf{\omega}=(\tau, 0, \varkappa)^T$ is the strain (Darboux) vector
in the Frenet frame, (b) the `constitutive' equations
\begin{equation}
M_b = {\cal E}_\varkappa(f), \quad\quad M_t = {\cal E}_\tau(f)
\label{eq:const_rel}
\end{equation}
and (c) the equations
\begin{equation}
{\cal E}_{\chi_i}(f) = 0, \quad\quad i = 1,2,\ldots,n,
\label{eq:phi}
\end{equation}
with ${\cal E}_\zeta$ the Euler-Lagrange operator for the variable
$\zeta$ defined by
${\cal E}_\zeta(h) = \partial_\zeta h -(\partial_{\zeta'}h)'+(\partial_{\zeta''}h)''-\ldots$.

{\it Note.}
We adopt the notation that for any vector $\bm{v} \in \mathbb{R}^3$ the
triple of components
$(v_t,v_n,v_b)=(\bm{v}\cdot\bm{t},\bm{v}\cdot\bm{n},\bm{v}\cdot\bm{b})$
will be denoted by the sans-serif symbol $\mathsf{v}$.
Equations (\ref{eq:balance}) in vectorial form read
$\bm{F}'=\bm{0}$, $\bm{M}'+\bm{r}'\times\bm{F}=\bm{0}$, the familiar balance
equations for a one-dimensional elastic continuum \citep{Antman05}. It
follows that $\bm{F}$ and $\bm{M}+\bm{r}\times\bm{F}$ are constant vectors in
space and that $|\mathsf{F}|$ and $\mathsf{F}\cdot\mathsf{M}$ are first
integrals.

{\it Proof.} It was proven by \citet{Anderson89} by performing the variation
of the curve that the Euler-Lagrange equations for $\varkappa$ and $\tau$ for
the problem in Eq.~(\ref{eq:funcl}) are given by
\iftwocolumns
\begin{eqnarray}
\varkappa {\cal H} + (\varkappa^2-\tau^2) {\cal E}_\varkappa + {\cal E}''_\varkappa+ \nonumber \\
+2\varkappa \tau {\cal E}_\tau +
\left(\frac{\varkappa\tau'-2\tau\varkappa'}{\varkappa^2}\right) {\cal E}'_\tau +2 \frac{\tau}{\varkappa}{\cal E}''_\tau =0, \label{eq:Anderson1} \\
\tau'{\cal E}_\varkappa + 2 \tau {\cal E}'_\varkappa - \varkappa'{\cal E}_\tau + \nonumber \\
+\left(\frac{\varkappa^2(\tau^2-\varkappa^2)-2{\varkappa'}^2+\varkappa\varkappa''}{\varkappa^3} \right){\cal E}'_\tau+ \nonumber \\
+2 \frac{\varkappa'}{\varkappa^2}{\cal E}''_\tau - \frac{1}{\varkappa} {\cal E}'''_\tau =0,
\label{eq:Anderson2}
\end{eqnarray}
\else
\begin{eqnarray}
\varkappa {\cal H} + (\varkappa^2-\tau^2) {\cal E}_\varkappa + {\cal E}''_\varkappa+ 
2\varkappa \tau {\cal E}_\tau +
\left(\frac{\varkappa\tau'-2\tau\varkappa'}{\varkappa^2}\right) {\cal E}'_\tau +2 \frac{\tau}{\varkappa}{\cal E}''_\tau =0, \label{eq:Anderson1} \\
\tau'{\cal E}_\varkappa + 2 \tau {\cal E}'_\varkappa - \varkappa'{\cal E}_\tau + 
\left(\frac{\varkappa^2(\tau^2-\varkappa^2)-2{\varkappa'}^2+\varkappa\varkappa''}{\varkappa^3} \right){\cal E}'_\tau+ 
2 \frac{\varkappa'}{\varkappa^2}{\cal E}''_\tau - \frac{1}{\varkappa} {\cal E}'''_\tau =0,
\label{eq:Anderson2}
\end{eqnarray}
\fi
where ${\cal H}={\cal H}(f)$ is the Hamiltonian
\iftwocolumns
\begin{eqnarray}
{\cal H}(f) = -f + \sum_{p\geq i>j\geq 0}\varkappa^{(i-j)}(-1)^j\frac{\mbox{d}^j}{\mbox{d}s^j}\left(\frac{\partial f}{\partial\varkappa^{(i)}}\right) + \nonumber \\
+\sum_{q\geq i>j\geq 0}\tau^{(i-j)}(-1)^j\frac{\mbox{d}^j}{\mbox{d}s^j}\left(\frac{\partial f}{\partial\tau^{(i)}}\right) + \nonumber \\
+\sum_{k=1}^{n} \sum_{r\geq i>j\geq 0}{\chi_k}^{(i-j)}(-1)^j\frac{\mbox{d}^j}{\mbox{d}s^j}\left(\frac{\partial f}{\partial{\chi_{k}}^{(i)}}\right). \label{eq:Ham}
\end{eqnarray}
\else
\begin{eqnarray}
{\cal H}(f) = -f + \sum_{p\geq i>j\geq 0}\varkappa^{(i-j)}(-1)^j\frac{\mbox{d}^j}{\mbox{d}s^j}\left(\frac{\partial f}{\partial\varkappa^{(i)}}\right) + \nonumber \\
+\sum_{q\geq i>j\geq 0}\tau^{(i-j)}(-1)^j\frac{\mbox{d}^j}{\mbox{d}s^j}\left(\frac{\partial f}{\partial\tau^{(i)}}\right) + 
\sum_{k=1}^{n} \sum_{r\geq i>j\geq 0}{\chi_k}^{(i-j)}(-1)^j\frac{\mbox{d}^j}{\mbox{d}s^j}\left(\frac{\partial f}{\partial{\chi_{k}}^{(i)}}\right). \label{eq:Ham}
\end{eqnarray}
\fi

Equations (\ref{eq:phi}) are nothing but the standard Euler-Lagrange
equations for the functions $\chi_k$. We now show that
Eqs.~(\ref{eq:balance}), (\ref{eq:const_rel}) are simply a rearrangement of
Eqs.~(\ref{eq:Anderson1}) and (\ref{eq:Anderson2}). Consider first the
equation for the moment and rewrite it in component form:
\begin{eqnarray}
M'_t -\varkappa M_n = 0, \label{eq:M1} \\
M'_n - \tau M_b + \varkappa M_t = -F_b,  \label{eq:M2} \\
M'_b + \tau M_n = F_n. \label{eq:M3}
\end{eqnarray}
Equation (\ref{eq:M1}) with the help of the second equation in
Eq.~(\ref{eq:const_rel}) allows us to express the principal normal component
as
\begin{equation}
M_n = {\cal E}'_\tau/\varkappa,
\label{eq:Mn}
\end{equation}
which we insert into Eqs.~(\ref{eq:M2}) and (\ref{eq:M3}) to find
\begin{eqnarray}
F_n = {\cal E}'_\varkappa + \frac{\tau}{\varkappa}{\cal E}'_\tau,  \label{eq:Fn} \\
F_b = \tau {\cal E}_\varkappa - \varkappa {\cal E}_\tau - \left(\frac{{\cal E}'_\tau}{\varkappa}\right)' . \label{eq:Fb}
\end{eqnarray}

Next we turn to the force equation (\ref{eq:balance}), which in component form reads
\begin{eqnarray}
F'_t -\varkappa F_n = 0, \label{eq:F1} \\
F'_n - \tau F_b + \varkappa F_t = 0,  \label{eq:F2} \\
F'_b + \tau F_n = 0. \label{eq:F3}
\end{eqnarray}
Now, it follows directly from Eq.~(\ref{eq:Ham}) that
\begin{equation}
{\cal H}' = -\varkappa' {\cal E}_\varkappa -\tau' {\cal E}_\tau
\label{eq:dHam}
\end{equation}
(here we have used that ${\cal E}_\chi =0$, by Eq.~(\ref{eq:phi})), and if we combine Eq.~(\ref{eq:dHam}) with Eqs.~(\ref{eq:F1}) and (\ref{eq:Fn}), and integrate, we obtain
\begin{equation}
F_t = {\cal H} + \varkappa {\cal E}_\varkappa + \tau {\cal E}_\tau + \mbox{const.}
\label{eq:Ft}
\end{equation}
The integration constant is fixed by the boundary conditions through the
integral $|\mathsf{F}|$ and can be absorbed into the Hamiltonian ${\cal H}$.
This defines all the force and moment components, and the two equations that
have not been used yet, Eqs.~(\ref{eq:F2}) and (\ref{eq:F3}), after
substitution of the force components from Eqs.~(\ref{eq:Fn}), (\ref{eq:Fb})
and (\ref{eq:Ft}), yield Eqs.~(\ref{eq:Anderson1}) and (\ref{eq:Anderson2}).

It is clear that the above steps can be carried out in the opposite direction,
i.e., by formally introducing new variables $F_t$, $F_n$, $F_b$, $M_t$, $M_n$,
$M_b$ according to the above expressions one can write
Eqs.~(\ref{eq:Anderson1}) and (\ref{eq:Anderson2}) as a first-order system.
Therefore, Eqs.~(\ref{eq:balance}) and (\ref{eq:const_rel}) are equivalent to
Eqs.~(\ref{eq:Anderson1}) and (\ref{eq:Anderson2}).
\hfill $\Box$

\smallskip
A few remarks are in order:
\begin{itemize}

\item[(i)]
Equations (\ref{eq:balance}) and (\ref{eq:const_rel}) can be
thought of as arising in two steps. In the first step $f$ is viewed as a
function of independent variables $\varkappa$ and $\tau$, and
Eqs.~(\ref{eq:const_rel}) are the classical Euler-Lagrange equations with
$M_b$ and $M_t$ playing the role of generalised forces. The order of
derivatives in the operators ${\cal E}_\varkappa$ and ${\cal E}_\tau$ is
determined by the order of derivatives of $\varkappa$ and $\tau$ appearing
in $f$. The second step then is to realise that $\varkappa$ and $\tau$ are
not arbitrary variables, but in fact the curvature and torsion of a space
curve. Eqs.~(\ref{eq:balance}), or equivalently Eqs.~(\ref{eq:Anderson1})
and (\ref{eq:Anderson2}), are then the result of expressing the variations
of $\varkappa$ and $\tau$ in terms of variations of the curve $\bm{r}$.
Since curvature is expressed as the second derivative of $\bm{r}$ and
torsion as the third derivative of $\bm{r}$, Anderson's equations involve
derivatives up to order two in ${\cal E}_\varkappa$ and up to order three
in ${\cal E}_\tau$. The balance equations (\ref{eq:balance}) are a rewrite
of these equations as a first-order system. The components of $\mathsf{M}$
couple the equations of step one to those of step two.

\item[(ii)]
The reason for calling Eqs.~(\ref{eq:const_rel}) `constitutive' equations is
that it is these equations that contain the physics of the problem (the
balance equations (\ref{eq:balance}) do not depend on $f$ explicitly).
Writing the Euler-Lagrange equations in the form of Eqs.~(\ref{eq:balance}),
(\ref{eq:const_rel}) and (\ref{eq:phi}) is a way of extracting constitutive
equations from the functional $f$. Mathematically, Eqs.~(\ref{eq:const_rel})
are best viewed as equations for $\varkappa$ and $\tau$, although they need
not be resolved for the highest derivatives of these variables.

\item[(iii)]
Equivalents of Eqs.~(\ref{eq:Anderson1}) and (\ref{eq:Anderson2}) have been
derived many times in the literature for particular applications. Examples
include the isotropic Kirchhoff rod \citep{Langer96}, the Helfrich rod
\citep{Wei98} (corrected in \cite{Liu06}), piezoelectric nanobelts
\cite{Tu06}, magnetic vortex filaments \cite{Barros07}, functionals that
involve either curvature or torsion or both \citep{Capovilla02}, a functional
that depends on curvature only \citep{Feoli05}, the Sadowsky functional for a
narrow developable strip \citep{Hangan05}, a functional that depends on
$\varkappa$, $\tau$ and their first derivatives \citep{Thamwattana08}, a
functional that involves $\varkappa$, $\tau$, $\chi_1$ and $\chi_1'$
\citep{Tu08}, etc. However, the explosion of terms that occurs when
${\cal E}_\varkappa$ and ${\cal E}_\tau$ are substituted makes
Eqs.~(\ref{eq:Anderson1}) and (\ref{eq:Anderson2}) not particularly practical
either for analytical or numerical study (for all but the very simplest
functionals $f$).

\item[(iv)]
It may happen that the right-hand sides in Eq.~(\ref{eq:const_rel})
have a simpler form in some other variables and accordingly we may prefer to
rewrite Eqs.~(\ref{eq:const_rel}) (and Eqs.~(\ref{eq:phi})) in these new
terms. Let the transformation be given by
\iftwocolumns
\begin{eqnarray}
\xi=\xi(\varkappa,\tau,\chi,\varkappa',\tau',\chi',\varkappa'',\tau'',\chi'',\ldots), \nonumber \\
\eta=\eta(\varkappa,\tau,\chi,\varkappa',\tau',\chi',\varkappa'',\tau'',\chi'',\ldots), \nonumber \\
\xi^{(i)} = \frac{\mbox{d}^{(i)}\xi}{\mbox{d}s^{(i)}}, \quad
\eta^{(i)} = \frac{\mbox{d}^{(i)}\eta}{\mbox{d}s^{(i)}}, \quad i=1,2,\ldots .\nonumber
\end{eqnarray}
\else
\begin{eqnarray}
\xi=\xi(\varkappa,\tau,\chi,\varkappa',\tau',\chi',\varkappa'',\tau'',\chi'',\ldots), \quad\quad
\xi^{(i)} = \frac{\mbox{d}^{(i)}\xi}{\mbox{d}s^{(i)}}, \quad i=1,2,\ldots ,\nonumber \\
\eta=\eta(\varkappa,\tau,\chi,\varkappa',\tau',\chi',\varkappa'',\tau'',\chi'',\ldots), \quad\quad
\eta^{(i)} = \frac{\mbox{d}^{(i)}\eta}{\mbox{d}s^{(i)}}, \quad i=1,2,\ldots .\nonumber
\end{eqnarray}
\fi
Then the Euler-Lagrange operators are transformed by~\cite{Damour91}
\begin{eqnarray}
{\cal E}_\varkappa(f)=\sum_{i=0}^\infty (-1)^i \frac{\mbox{d}^i}{\mbox{d}s^i}\left(\frac{\partial \xi}{\partial{\varkappa}^{(i)}}{\cal E}_\xi(\tilde{f})+\frac{\partial \eta}{\partial{\varkappa}^{(i)}}{\cal E}_\eta(\tilde{f})
\right), \label{eq:trans_k} \\
{\cal E}_\tau(f)=\sum_{i=0}^\infty (-1)^i \frac{\mbox{d}^i}{\mbox{d}s^i}\left(\frac{\partial \xi}{\partial{\tau}^{(i)}}{\cal E}_\xi(\tilde{f})+\frac{\partial \eta}{\partial{\tau}^{(i)}}{\cal E}_\eta(\tilde{f})
\right) , \label{eq:trans_t}
\end{eqnarray}
where $\tilde{f}$ is the transformed $f$
(similar expressions hold for $\chi$ as in the usual case of Lagrangians involving higher-order derivatives).
\end{itemize}

\section{Examples}
We illustrate the above theory by several examples.

{\it The anisotropic Kirchhoff rod}~\cite{Langer96}. Let the curvature
$\varkappa(s)$ and torsion $\tau(s)$ define the centreline $\bm{r}(s)$ of the
rod (up to Euclidean motions). Assuming a non-circular cross-section with
bending stiffnesses $A$ and $B$ and torsional stiffness $C$, we can write
the elastic energy density as~\cite{Zhao06}
\begin{equation}
f(\varkappa,\tau,\phi,\phi') = (a+b\cos{2\phi})\varkappa^2 + c (\tau + \phi')^2,
\label{eq:Kirod}
\end{equation}
where $a=(A+B)/4$, $b=(B-A)/4$, $c=C/2$ and $\phi$ is the twist angle
describing the rotation of the local material frame with respect to the
Frenet frame about the tangent vector $\bm{t}=\bm{r}'$.
With $\phi$ playing the role of $\chi_1$, Eqs.~(\ref{eq:const_rel}) and
(\ref{eq:phi}) then give, respectively,
\begin{eqnarray}
M_b=\partial_\varkappa f = 2(a+b\cos{2\phi})\varkappa, \label{eq:Mb_kir}\\
M_t=\partial_\tau f = C(\tau+\phi') \label{eq:Mt_kir}
\end{eqnarray}
and
\begin{equation}
c(\tau'+\phi'')+b\varkappa^2\sin{2\phi} = 0 .
\label{eq:phi_kir}
\end{equation}
Equations (\ref{eq:balance}), (\ref{eq:Mb_kir}), (\ref{eq:Mt_kir},
(\ref{eq:phi_kir}) constitute a system of differential-algebraic equations
(DAEs) that can be turned into a system of ODEs by differentiation of the
algebraic equations.
For an isotropic rod ($A=B$) the coefficient $b$ vanishes and a
combination of Eq.~(\ref{eq:phi_kir}) and Eq.~(\ref{eq:Mt_kir}) gives the
first integral $M_t=:\bar{c}=\const$, which allows the system to be
integrated in closed form. In this case the equation for the angle
$\phi$ fully decouples from the other equations and the centreline of the
isotropic rod can be found as a minimiser of the functional
$f=a\varkappa^2+\bar{c}\tau$ with a linear torsion term \cite{Langer96}.
On the other hand, the functional $f=\frac{1}{2}A\varkappa^2+\frac{1}{2}C\tau^2$ with quadratic
torsion was proposed to model elastic strips and polymer
chains~\cite{Mahadevan93,Kessler03}. It may be formally obtained from
Eq.~(\ref{eq:Kirod}) by pushing one of the bending stiffnesses, $B$, to
infinity (implying $\phi\to \pi/2$).
Rods described by this functional bend only about a single principal axis and
therefore have their material frame locked to the Frenet frame.

For a bundle of parallel thin rods of circular cross-section of radius $R$
the normalised bending energy density may be shown to equal \cite{Starostin06}
\begin{equation}
f=\left(1-\sqrt{1-R^2\varkappa^2}\right),
\label{eq:bundle}
\end{equation}
which provides another example of an invariant functional \endnote{Note the
similarity between Eq.~(\ref{eq:bundle}) and the Lagrangian for a relativistic
particle of maximal proper acceleration (Eq.~(4.4) in \cite{Nesterenko95})
though the ambient spaces are different.}.
Ref.~\cite{Grason09} gives
extensions to more complicated functionals for parallel bundles to which our
proposition can be applied to derive equilibrium equations.

{\it The Helfrich rod.} In order to study chiral effects in polymers Helfrich
proposed the following elastic energy density with higher-order terms
included~\cite{Helfrich91}
\begin{equation}
f=f(\varkappa,\tau,\varkappa') = \frac{k_2}{2}\varkappa^2 +k_3\varkappa^2\tau +\frac{k_{22}}{4}\varkappa^4 + 
\frac{k_4}{2} (\varkappa'^2+\varkappa^2\tau^2) ,
\label{eq:Helfrich}
\end{equation} 
where $k_2, k_3, k_{22}, k_4$ are constant coefficients. For this functional
Eqs.~(\ref{eq:const_rel}) become
\begin{eqnarray}
M_b=k_2\varkappa + 2 k_3 \varkappa\tau +k_{22}\varkappa^3+k_4\varkappa\tau^2-k_4\varkappa'', \label{eq:Mb_hlf}\\
M_t=k_3\varkappa^2+k_4\varkappa^2\tau  \label{eq:Mt_hlf}.
\end{eqnarray}
These are the nonlinear constitutive equations for the Helfrich rod (expressed
in the Frenet frame). The second equation is algebraic and can be used to
eliminate the torsion $\tau$. The first equation is then a differential
equation for $\varkappa$ that is to be solved in conjunction with the balance
equations.

The Helfrich functional has been extended to sixth order, involving the first
derivative of torsion and the second derivative of curvature \cite{Liu03}.

{\it A rod lying in a surface.}
The proposition can also be used in problems of curves with constraints such
as the constraint for a rod to lie in a surface. If this surface constraint
is given by the pointwise condition
$0=g(\varkappa,\tau,\psi,\varkappa',\tau',\psi',\ldots)\in \mathbb{R}^m$,
$\psi \in \mathbb{R}^{m-1}$, for certain $m$, then we consider the new
functional $f+\lambda(s)\cdot g$ with $\lambda(s)\in \mathbb{R}^m$ a Lagrange
multiplier.

The simplest example is that of a rod in a {\it plane}.
One may constrain the centreline $\bm{r}=(x,y,z)$ to a plane by imposing,
for instance, $z=0$, as in \cite{Heijden99}, but this constraint is not
Euclidean invariant and therefore not of the type $g$ above. A Euclidean
invariant form is simply $\tau=0$. We can account for this constraint by
modifying the function in Eq.~(\ref{eq:Kirod}) and considering
$f_1=f+\lambda(s)\tau$ (hence $\chi_1=\phi$, $\chi_2=\lambda$).
Equation~(\ref{eq:Mb_kir}) does not change while Eq.~(\ref{eq:Mt_kir})
now becomes $M_t= C(\tau+\phi')+\lambda$ and may be used to find the reaction
$\lambda$. The binormal force component is constant by virtue of
Eq.~(\ref{eq:F3}). The remaining five Eqs.~(\ref{eq:F1}), (\ref{eq:F2}),
(\ref{eq:M1}), (\ref{eq:M2}), (\ref{eq:M3}) plus Eq.~(\ref{eq:phi_kir}) with
$\tau\equiv 0$ and $\varkappa$ substituted from Eq.~(\ref{eq:Mb_kir}) form a
system of six differential equations for the six variables
$M_t, M_n, M_b, F_t, F_n, \phi$.

Note that the reaction $\lambda$ has the interpretation of a moment about the
tangential direction. The constraint may therefore be realised by applying
a distributed twisting couple of the same magnitude. It may be approximated
by a rod with multiple (in the limit -- distributed) small whiskers
perpendicular to the centreline (not unlike a caterpillar).
If we imagine placing such a hairy rod between two parallel friction-free
plates so that the rod itself would not get in touch with the plates then
the normal reaction forces would give the required couples.

This way of realising the constraint differs of course from the usual one
corresponding to the $z=0$ condition, where the reactions are distributed
normal {\it forces} exerted by the plates onto the rod \cite{Heijden99}.
Anyway, if one is only interested in the configuration of the rod then the
realisation of the constraint does not matter. In particular, for the
isotropic rod the equations in both cases reduce to those of the Euler
elastica, $\varkappa''+\frac{1}{2}\varkappa^3=0$, corresponding to the
functional $f=\varkappa^2$.

The Helfrich rod can be similarly constrained to the plane by introducing the
condition $\tau=0$. A more direct way to obtain the reduced functional is to
delete the torsion terms in the right
hand side of Eq.~(\ref{eq:Helfrich}) to obtain
$f= \frac12 k_2\varkappa^2  +\frac14 k_{22}
\varkappa^4 + \frac12 k_4 \varkappa'^2$ and consider the problem in
$\mathbb{R}^2$. This functional may be useful for studying polymers
synthesised at the interface of two fluids.
It may also have application in computer vision. In this field the functional
$f=\varkappa'^2$ has been proposed for shape completion \cite{Kimia03}.
The Euler-Lagrange equation for this functional,
$\varkappa'''+\varkappa^2\varkappa''-\frac{1}{2}\varkappa\varkappa'^2=0$,
follows directly from Anderson's Eq.~(\ref{eq:Anderson1}) \cite{Anderson89}
(the equation in \cite{Kimia03} is incorrect).
Functionals that involve the torsion may be of interest when one deals with
completion of space curves reconstructed from their planar projections.

Rods confined to a {\it cylinder} are relevant for buckling inside tubes
and for supercoiled filaments, and have been studied by imposing the
coordinate constraint $x^2+y^2=R^2$, where $R$ is the radius of the cylinder
\cite{Heijden01}. A Euclidean invariant form of the constraint involves two
conditions~\cite{Ko67}:
\begin{eqnarray}
g_1 := \varkappa^2 -\varkappa_0^2 \cos^4\theta - \theta'^2 = 0 , \nonumber \\
g_2 := \varkappa\theta'' -\varkappa'\theta'+\varkappa_0 \varkappa\cos^2\theta(\varkappa_0 \sin\theta\cos\theta-\tau)=0,
\end{eqnarray}
where $\varkappa_0^{-1}=R$ and $\theta(s)$ is an unknown function that is to
be found as part of the solution.
The modified functional $f+\lambda_1(s)g_1+\lambda_2(s)g_2$ is of the required
form in Eq.~(\ref{eq:funcl}) (with $\chi_1 = \lambda_1, \chi_2 = \lambda_2,
\chi_3 = \theta$) and the Euler-Lagrange equations follow from the
proposition.

{\it Inextensible strips.}
An inextensible strip is a thin shell that deforms by pure bending (no
stretching). Its surface is therefore developable and has a single non-zero
principal curvature $\varkappa_1$. The normalised bending energy for a
rectangular strip of length $L$ and width $2w$ can be reduced to a single
integral over the strip's centreline \cite{Wunderlich62}:
\begin{eqnarray}
\int_0^L\int_{-w}^w\varkappa_1^2(s,t)\,\mbox{d}t \ \mbox{d}s = \int_0^L f(\varkappa,\eta,\eta')\,\mbox{d}s, \nonumber \\
f(\varkappa,\eta,\eta')=\varkappa^2\left(1+\eta^2\right)^2\frac{1}{w\eta'}
\log\left(\frac{1+w\eta'}{1-w\eta'}\right),
\label{eq:g_function}
\end{eqnarray}
where $\eta=\tau/\varkappa$. In the limit $w\to 0$ this recovers Sadowsky's
functional $f(\varkappa,\eta)=2\varkappa^2\left(1+\eta^2\right)^2$ given in
\cite{Sadowsky31} where Eqs.~(\ref{eq:const_rel}) for this case are obtained
by applying the principle of virtual work and making use of the variation of
the Frenet frame.
Since the energy density $f$ depends on derivatives of the curvature only via
$\eta$, it is convenient to apply the transformation $\xi=\varkappa$,
$\eta=\tau/\varkappa$. Eqs.~(\ref{eq:trans_k}), (\ref{eq:trans_t}) then yield
$M_b={\cal E}_\varkappa(f)-\frac{\eta}{\varkappa}{\cal E}_\eta(f)$,
$M_t=\frac{1}{\varkappa}{\cal E}_\eta(f)$.
Note that ${\cal E}_\varkappa(f) = \partial_\varkappa f$ and hence $M_b+\eta M_t = \partial_\varkappa f$.
These equations were first derived in \cite{Starostin07}.
The complexity of the centreline-reduced functional $f(\varkappa,\eta,\eta')$
makes this the first problem for which the invariant formulation seems to be
the only way to obtain a manageable set of equilibrium equations. Their
extension to intrinsically curved strips was considered in
\cite{Starostin08a}.

The balance equations presented here correspond to the conservation laws
generated by the symmetry group of Euclidean motions \cite{Capovilla02}.
A computational procedure for deriving invariant Euler-Lagrange equations
(analogous to Eqs.~(\ref{eq:Anderson1}) and (\ref{eq:Anderson2})) for
arbitrary finite-dimensional transformation groups can be found in
\cite{Kogan03}. When given the balance form these equations may be useful
for certain problems with non-Euclidean symmetry groups. An example is the
description of world lines of relativistic particles in Minkowski space
with the Poincar\'e group of isometries as symmetry group
\cite{Nesterenko95,Ferrandez06}.

\bibliography{star}

\end{document}